# Fishtail effect and the vortex phase diagram of high-entropy alloy superconductor

Lingling Gao[1,2], Tianping Ying[3], Yi Zhao[1], Weizheng Cao[1], Changhua Li[1], Lin Xiong[1], Qi Wang[1,4], Cuiying Pei[1], Jun-Yi Ge[5], Hideo Hosono[6] and Yanpeng Qi[1,4*]

1. School of Physical Science and Technology, ShanghaiTech University, Shanghai 201210, China
2. University of Chinese Academy of Science, Beijing 100049, China
3. Beijing National Laboratory for Condensed Matter Physics, Institute of Physics, Chinese Academy of Sciences, Beijing 100190, China
4. ShanghaiTech Laboratory for Topological Physics, ShanghaiTech University, Shanghai 201210, China
5. Materials Genome Institute, Shanghai University, Shanghai 200444, China
6. Materials Research Center for Element Strategy, Tokyo Institute of Technology, Yokohama 226-8503, Japan

**ABSTRACT**

**High-entropy alloy (HEA) is an attracting topic raising in materials science and condensed matter physics. Although several types of superconductors have been discovered in HEAs, the critical currents ($J_c$) of HEA superconductors remain uncharacterized up to now. Here, we systematically study the current-carrying ability of $(TaNb)_{0.7}(HfZrTi)_{0.5}$ HEA at various heat treatment conditions. We obtained the high upper critical field and large current carrying ability, which point to optimistic applications. Interestingly, the fishtail or second peak effect was found for the first time in HEA superconductors, and the position of the vortex pinning force shows a maximum at 0.72 of the reduced field, which is quite different from the cuprates and iron-based high-$T_c$ superconductors. Together with the resistive measurements, the vortex phase diagram is obtained for HEA superconductor.**

* Correspondence should be addressed to Y.P.Q. (qiyp@shanghaitech.edu.cn)

High-entropy alloys (HEAs), a new class of materials that proposed in 2004[1, 2], are typically composed of five or more major elements in similar concentrations, ranging from 5 to 35 at. % for each element. The atoms randomly distribute on simple crystallographic lattices, where the high entropy of mixing can stabilize disordered solid-solution phases with simple structures[3]. In addition to their structural and chemical diversity, HEAs can display novel, highly tunable properties, for example, excellent specific strength[4, 5], superior mechanical performance at high temperatures [6, 7], and fracture toughness at cryogenic temperatures[8, 9], making them promising candidates for new applications. Since their properties often fall between crystalline and amorphous states, HEAs are good model systems for investigating fundamental physical interactions. In 2014, the first HEA superconductor was found in Ta-Nb-Hf-Zr-Ti system composed of $4d$ and $5d$ series elements[10]. Until now, four types of superconductors have been discovered in HEAs, which are the body-centered cubic (bcc)[11-14], CsCl-type[15], α-Mn-type[16], and hexagonal-closest packing (hcp)[17] crystal structures. Moreover, the superconductors of "HEA-type" have been synthesized[18, 19].

HEA superconductor displays not only excellent mechanical properties[20], but also robust superconductivity[21, 22] and quiet high upper critical field[3], which appear to be favorable for potential practical applications. The flux-pinning mechanism which governs the field and temperature dependence of critical current density is very important to the practical application. However, the critical currents ($J_c$) of HEA superconductors remain uncharacterized up to now. In this work, we performed magnetotransport and magnetization measurements to study the current carrying ability of HEA superconductors. We focused $(TaNb)_{0.7}(HfZrTi)_{0.5}$, which belong to bcc-type HEA superconductors. The fishtail or second peak effect was found first time in HEA superconductors, and the position of the vortex pinning force shows a maximum at 0.72 of the reduced field, which is quite different from the cuprates and iron-based high-$T_c$ superconductors. We obtained the high upper critical field $H_{c2}$ (10 T) and large current carrying ability $J_c$ ($10^5$ A/cm$^2$), which point to optimistic applications. Together with the resistivity measurements, for the first time the vortex phase diagram is obtained for HEA superconductors.

A polycrystalline sample of $(TaNb)_{0.7}(HfZrTi)_{0.5}$ was synthesized by arc melting method[12]. The pure metals were arc-melted using high currents (at $T > 2500$ °C) in an argon atmosphere. The samples were flipped and remelted several times with negligible mass loss to ensure the homogeneity of the ingot. Finally, the alloy was rapidly cooled on a water-chilled copper plate. The sample was cut to cuboid-shape, sealed in quartz ampules under partial argon atmosphere, and annealed at corresponding $T_a$ for 5 h. X-ray diffraction (XRD) pattern was carried out at room temperature by a Bruker D8 Advance with Cu $K_α$ radiation ($λ = 0.15418$ nm). The morphology and composition of samples were observed by scanning electron microscopy (SEM) with energy dispersive spectroscopy (EDS). For measuring high-resolution transmission electron microscopy (HRTEM) images, the samples were fabricated into a rectangular shape using the focus ion beam (FIB) technique using JEOL JIB-4700F. The HRTEM images were captured using a JEM-F200. The electrical resistivity was measured using a standard four-probe method by Quantum Design PPMS-9T. Magnetization measurements were performed using Quantum Design MPMS3 and the $J_c$ obtained from MHLs by Bean model[23].

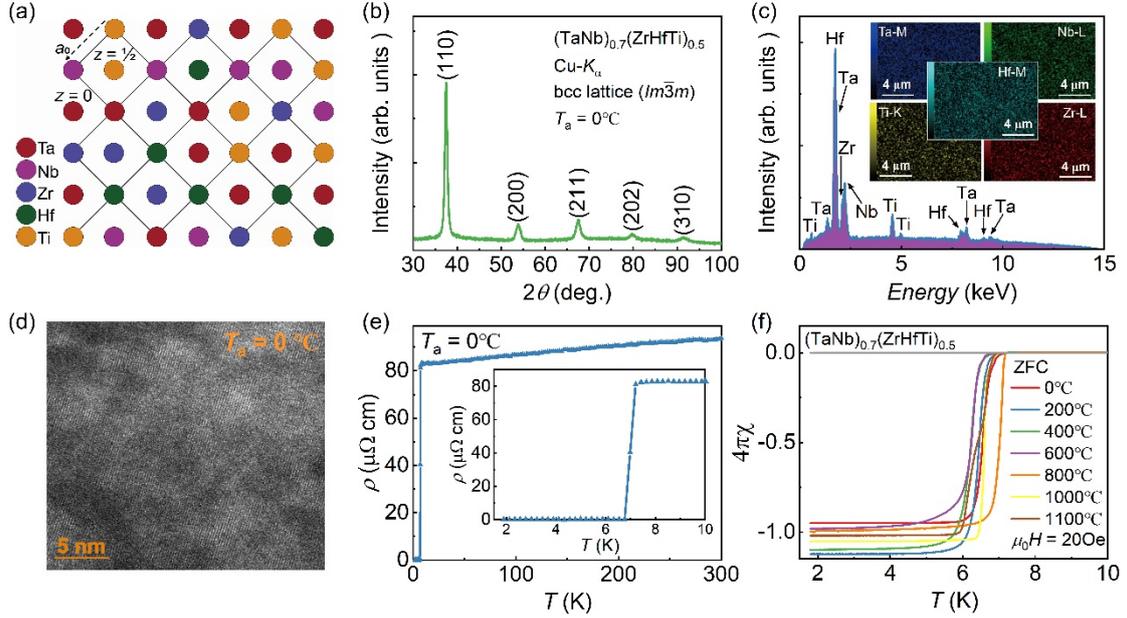

FIG. 1. (a) Illustration of the crystal structure of $(TaNb)_{0.7}(HfZrTi)_{0.5}$ HEA with bcc lattice. (b) XRD pattern of $(TaNb)_{0.7}(HfZrTi)_{0.5}$ without heat treatment ($T_a = 0$ °C) at room temperature. (c) The EDS spectrum of $(TaNb)_{0.7}(HfZrTi)_{0.5}$ without heat treatment. The insets show the EDS mapping of five elements, respectively. (d) Nanostructure of the $(TaNb)_{0.7}(HfZrTi)_{0.5}$ depicted in a HRTEM image. (e) Temperature dependence of resistivity $\rho(T)$ of $(TaNb)_{0.7}(HfZrTi)_{0.5}$ without heat treatment. Inset: Enlarged view of low-temperature region, showing superconducting transition. (f) DC magnetization data at low temperature under applied magnetic field of 20 Oe for various annealing temperature HEA.

$(TaNb)_{0.7}(HfZrTi)_{0.5}$ HEA crystallizes in bcc crystal structure with space group of $Im\bar{3}m$[3]. The atoms of $(TaNb)_{0.7}(HfZrTi)_{0.5}$ HEA randomly arrange themselves on the crystallographic positions [Fig. 1(a)], achieving high configurational entropy. As shown in Fig. 1(b), all XRD peaks can be indexed by bcc structure with $a = 3.405$ Å, which is excellent agreement with the theoretical value $a_{min} = 3.39$ Å[10]. It should be noted that the peaks are quite broad, revealing random mixing of the elements. This is further confirmed by the mapping of energy dispersive X-ray analysis, where the five elements distribute uniformly in large area scale [the inset of Fig. 1(c)]. Fig.1(d) shows a representative HRTEM image of $(TaNb)_{0.7}(HfZrTi)_{0.5}$ HEA and no nanoscale chemical phase separation was observed. In Fig. 1(e), the typical $\rho(T)$ curve for $(TaNb)_{0.7}(HfZrTi)_{0.5}$ is shown in the temperature range 1.8 -300 K. The $\rho(T)$ data exhibits a weak positive temperature coefficient with a room-temperature resistivity of 93.79 μΩ cm. A sharp drop is observed below 7.2 K [inset of Fig. 1(e)], indicating the onset of superconductivity. Further evidence for bulk superconductivity was obtained from the large diamagnetic signal at 7.13 K, shown in Fig. 1(f).

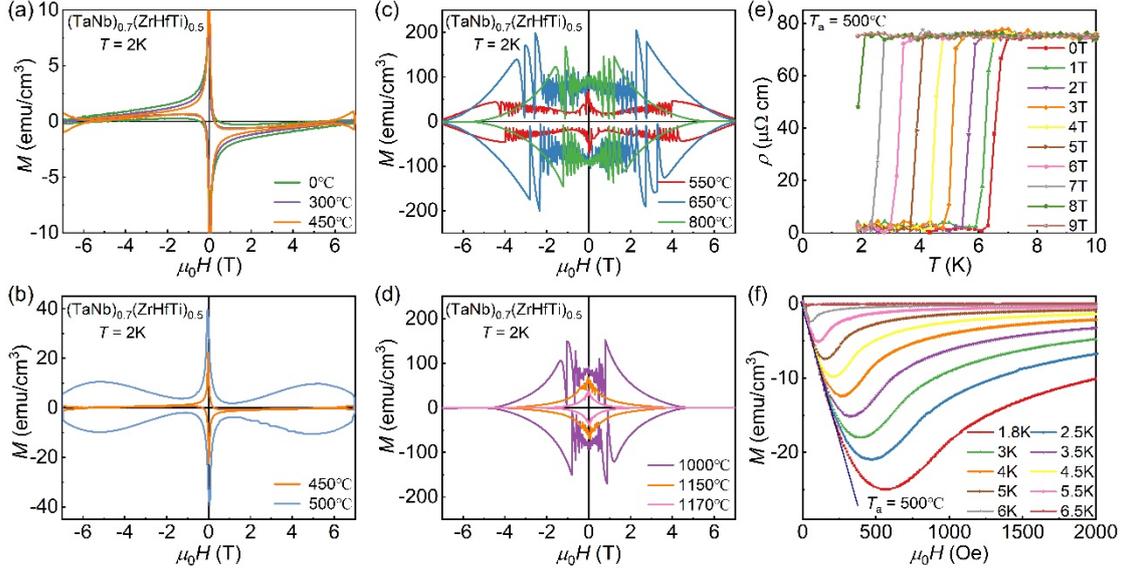

FIG. 2. Evolution of the magnetization as a function of annealing temperature ($T_a$) for $(TaNb)_{0.7}(HfZrTi)_{0.5}$. (a) and (b) Magnetization curves with the applied field at 2 K in the low annealing temperature. The fishtail effect is observed. (c) and (d) The flux jump appears and the region of flux jump gradually shrinks inward with increasing annealing temperature. (e) Temperature dependence of the resistivity at various external field for the sample after annealed at 500 °C. (f) Low-field parts of $M(H)$ at various temperatures of the sample annealed at 500 °C. The blue solid line is the Meissner line as discussed in the text.

In order to remove the stress, the sample was annealed at various temperatures. At the same time, we measured the magnetic hysteresis loops (MHLs) at 2 K for the sample annealing at different temperature. For the sample without annealing, we observe the typical MHL, which is same as normal alloy superconductor [Fig. 2(a)]. Interesting, a small "fish-tail" hump appears at high fields for the sample annealed at 450 °C. As shown in Fig. 2(b), the second peak can be easily observed for the sample annealed at 500 °C. This is the so-called fishtail effect or second peak effect[24, 25]. Although this feature has been observed in the clean and high quality single crystals of cuprate[24, 26, 27] and Fe-based superconductors[25, 28] as well as in the low $T_c$ superconductors[29-31], now this is the first time we observed in HEA superconductors. When further increasing the annealing temperature ($T_a$ > 550 °C), the flux jump occurs in the low field region and disappear until the $T_a$ up to 1170°C. Herein, we will focus the sample with fishtail effect ($T_a$ = 500 °C) to carry out detailed critical field and current carrying ability measurements.

Fig. 2(e) shows temperature dependence of the resistivity below 10 K at various magnetic fields. With increasing magnetic field, the superconducting transition gradually shifted to a lower temperature, while the superconducting transition width was almost unchanged. We plotted the upper critical field ($\mu_0H_{c2}(T)$) using 90% points on the resistivity transition curves as a function of temperature in Fig. S1(c) (supplementary material). The $\mu_0H_{c2}(T)$ are fitted by using Ginzburg-Landau (G-L) formula,

$$\mu_0H_{c2}(T) = \mu_0H_{c2}(0)\ (1-(T/T_c)^2)/(1+(T/T_c)^2) \tag{1}$$

The estimated $\mu_0H_{c2}(0)$ values is 10.07 T, which yields a Ginzburg–Landau coherence length $\xi_{GL}(0)$ of 5.72 nm. Fig. 2(f) show the low field magnetization curves $M(H)$ taken at different temperatures. The lower critical field $\mu_0H_{c1}(T)$ in Fig. S1(d) (supplementary material) is defined as the field deviating from the linearity of the initial slope in the magnetization. The linear extrapolation of $\mu_0H_{c1}(T)$ to $T = 0$ K show $\mu_0H_{c1}(0)$ to be 227.51 Oe. The superconducting critical parameters are similar to what we observed in the sample without annealing.

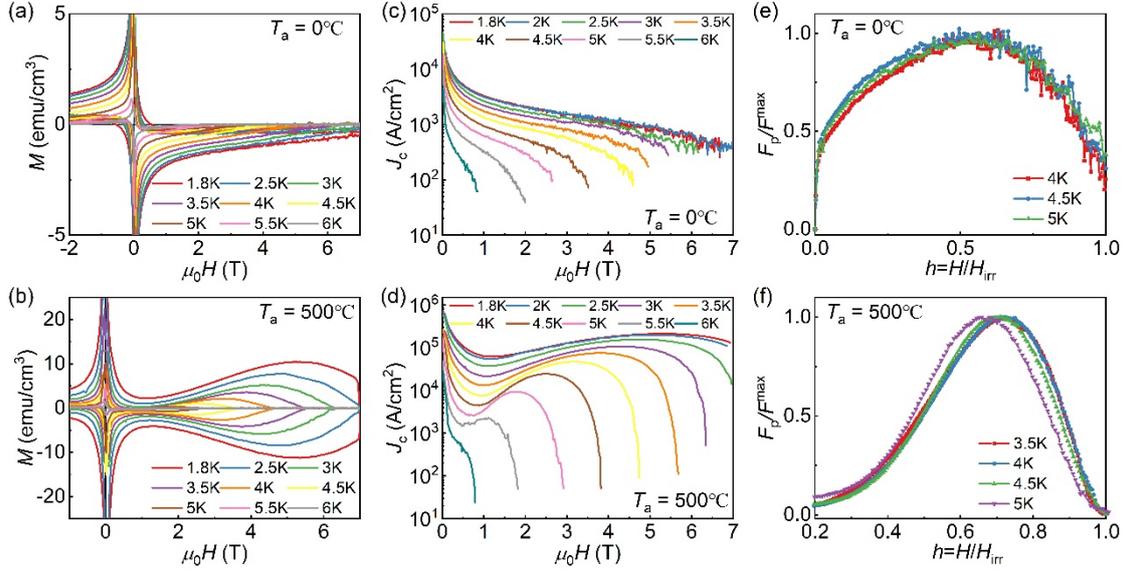

FIG. 3. (a) and (b) MHLs measured at different temperatures for the sample without heat treatment ($T_a = 0$ °C) and annealed at 500 °C, respectively. (c) and (d) Field dependence of the calculated critical current density from the Bean model at different temperatures for the sample without heat treatment ($T_a = 0$ °C) and annealed at 500 °C, respectively. Scaling behavior of pinning force for (e) the sample without heat treatment ($T_a = 0$ °C) and (f) annealed at 500 °C. The irreversibility field $H_{irr}$ is determined from $J_c$-$\mu_0H$.

In Figs. 3(a) and 3(b), the MHLs of the sample without heat treatment ($T_a = 0$ °C) and annealed at 500 °C measured at different temperatures from 1.8 to 6 K are presented respectively. For the annealing sample, the second peak can be easily observed and moves to the high field with the decrease in temperature, which is significant different from the sample without annealing. The width of the irreversible magnetization $\Delta M$ increases with increasing of the magnetic field, which shows an obvious fishtail effect. The critical current density $J_c$ calculated from the MHLs using the Bean model[23, 32, 33] are shown in Figs. 3(c) and 3(d). $J_c$ of annealing sample exhibits a broad maximum in the high field due to the second peak. The self-field $J_c = 6.81 \times 10^5$ A/cm$^2$ at 2 K, which is one order higher than that of sample without annealing. To assess mechanisms, which control the vortex pinning force $F_p \propto J_cH$, we plot in Figs. 3(e) and 3(f) the normalized pinning force $f = F_p / F_p^{max}$ as a function of the reduced field $h = H / H_{irr}$. The symmetric $f(h, T)$ curves of sample without annealing overlap well with a peak at $h \sim 0.5$, as shown in Fig. 3(e). According to the theoretical model[34, 35], the dominant pinning is body pinning, perhaps resulting from random distribution of atoms in HEA. For the sample annealed at 500 °C, the normalized curves of $f(h, T)$ at high temperatures near $T_c$ exhibit a peak at around $h \sim 0.72$, which is dramatically different from YBCO and $Ba_{0.6}K_{0.4}Fe_2As_2$ with $h \sim 0.33$[24, 25]. In order to figure out the pinning center, we performed the XRD measurements after annealing at various temperatures. As shown in Fig. S2 (supplementary material), bcc crystal structure of $(TaNb)_{0.7}(HfZrTi)_{0.5}$ HEA is stable until annealing at 500 °C. After annealing at 700 °C, a set of new diffraction peaks emerges, which could be attributed to TaNb with bcc structure (space group $Im\bar{3}m$). As shown in Fig. S3, the clusters were precipitated from HEA with size of ~ 10 nm (supplementary material). Therefore, the fishtail effect in the high temperature region of this sample is from the dense vortex pinning nanostructure, which is from the weak precipitation during the heat treatment process.

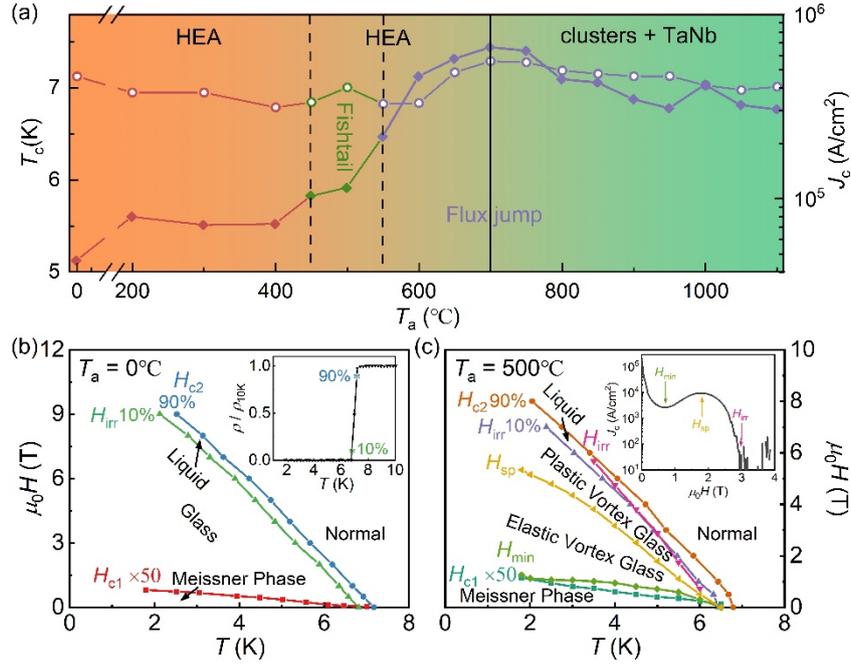

FIG. 4. (a) Annealing temperature dependence of superconducting transition temperatures and of the maximum critical current density at 2 K for $(TaNb)_{0.7}(HfZrTi)_{0.5}$ HEA. The open circle represents the $T_c$ obtained from the $\chi$-$T$ curve, and the solid square represents the maximum value of the $J_c$ at 2 K. There are total of four regions. The two different colored regions separated by a solid black line indicate different crystal structures. The green and purple lines respectively show the fishtail effect and flux jump of MHLs, which are parted by dotted lines. The vortex phase diagram of $(TaNb)_{0.7}(HfZrTi)_{0.5}$ HEA (b) without heat treatment ($T_a = 0$ °C) and (c) annealed at 500 °C. The inset of (b) shows the definitions of $H_{c2}$ and $H_{irr}$ from the resistivity measurement, and the inset of (c) shows the definitions of three characteristic fields from the magnetic measurement.

The $T_c$ and $J_c$ as function of $T_a$ are summarized in Fig. 4(a). After annealing, the $T_c$ remains nearly unchanged while $J_c$ enhancement more than one order due to induce vortex pinning center. The appearance of flux jump is associated with the low thermal conductivity of HEAs[36-38]. Annealing at higher temperature could reduce the flux jump accompanying by the structural phase transition. The vortex phase diagram for the sample without heat treatment ($T_a = 0$ °C) and after annealed at 500 °C are present in Figs. 4(b) and 4(c), respectively. The upper critical field $H_{c2}$ with 90 % $\rho_n$ and the irreversibility field $H_{irr}$ with 10% $\rho_n$ are shown as stars in the inset of Fig. 4(b). The vortex-liquid regime presents between the $H_{c2}$ line and the vortex-glass phase[39]. Three characteristic fields are confirmed as shown by the solid symbols in the inset of Fig. 4(c)[25]. $H_{min}$ and $H_{sp}$ locate at the valley and the peak of the curve respectively, and the irreversibility field $H_{irr}$ is also obtained from taking a criterion of 100 A/cm$^2$. The irreversibility lines determined by magnetic and resistive measurements are basically coincident for the sample annealed at 500 °C. The $H_{min}$-$T$ and $H_{sp}$-$T$ curves cannot be fitted by the expressions $H^*(T) = H^*(0) \times (1-T/T_c)^\alpha$, which is different from the cuprates and iron-based superconductors. The formation of the vortex phase diagram may relate to the disordered distribution of elements in the HEA. However, the real pinning mechanism needs further investigation.

In conclusion, the evolution of the critical current density in (TaNb)$_{0.7}$(HfZrTi)$_{0.5}$ HEA is investigated under various annealing temperatures. We find the fishtail effect as well as large current carrying ability in HEA superconductor. The fishtail effect at high temperatures below $T_c$ is originated from the small size normal core pinning effect. The vortex phase diagram of HEA superconductor is obtained for the first time. Consider the excellent mechanical properties and large upper critical field, all contributing to potential applications in HEA superconductors.

This work was supported by the National Key R&D Program of China (Grant No. 2018YFA0704300), the National Natural Science Foundation of China (Grant No. U1932217, 11974246, 12004252), Natural Science Foundation of Shanghai (Grant No. 19ZR1477300), and the Science and Technology Commission of Shanghai Municipality (No. 19JC1413900). The authors thank the support from CℏEM (No. 02161943) and Analytical Instrumentation Center (No. SPST-AIC10112914), SPST, ShanghaiTech University.

## DATA AVAILABILITY

The data that support the findings of this study are available from the corresponding authors upon reasonable request.